\documentclass[conference]{IEEEtran}
\IEEEoverridecommandlockouts
\usepackage{cite}
\usepackage{amsmath,amssymb,amsfonts}
\usepackage{algorithm}
\usepackage{textcomp}
\usepackage{mathtools}
\usepackage{algpseudocode}
\usepackage{graphicx}
\usepackage{textcomp}
\usepackage{xcolor}
\definecolor{airforceblue}{rgb}{0.36, 0.54, 0.66}
\definecolor{applegreen}{rgb}{0.55, 0.71, 0.0}
\definecolor{ao(english)}{rgb}{0.0, 0.5, 0.0}
\definecolor{ao}{rgb}{0.0, 0.0, 1.0}
\usepackage{caption}
\usepackage{subcaption}
\usepackage{enumitem}

\def\BibTeX{{\rm B\kern-.05em{\sc i\kern-.025em b}\kern-.08em
    T\kern-.1667em\lower.7ex\hbox{E}\kern-.125emX}}

\begin{document}
\thispagestyle{empty} This paper has been accepted for publication in the proceedings of the IEEE International Workshop on Computer Aided Modeling and Design of Communication Links and Networks (CAMAD).

\copyright 2023 IEEE. Personal use of this material is permitted. However, permission from IEEE must be obtained for all other uses, whether in current or future media. This includes reprinting/republishing this material for advertising or promotional purposes, creating new collective works, resale or redistribution to servers or lists, or the reuse of any copyrighted component of this work in other works.

\title{\huge Edge AI Inference in Heterogeneous Constrained Computing: Feasibility and Opportunities}

\author{\IEEEauthorblockN{Roberto Morabito}
\IEEEauthorblockA{\textit{University of Helsinki} \\
roberto.morabito@helsinki.fi}
\and
\IEEEauthorblockN{Mallik Tatipamula}
\IEEEauthorblockA{\textit{Ericsson} \\
mallik.tatipamula@ericsson.com}
\and
\IEEEauthorblockN{Sasu Tarkoma}
\IEEEauthorblockA{\textit{University of Helsinki} \\
sasu.tarkoma@helsinki.fi}
\and
\IEEEauthorblockN{Mung Chiang}
\IEEEauthorblockA{\textit{Purdue University} \\
chiang@purdue.edu}
}

\maketitle
\pagestyle{plain}
\begin{abstract}

The network edge's role in Artificial Intelligence (AI) inference processing is rapidly expanding, driven by a plethora of applications seeking computational advantages. These applications strive for data-driven efficiency, leveraging robust AI capabilities and prioritizing real-time responsiveness. However, as demand grows, so does system complexity. The proliferation of AI inference accelerators showcases innovation but also underscores challenges, particularly the varied software and hardware configurations of these devices. This diversity, while advantageous for certain tasks, introduces hurdles in device integration and coordination. In this paper, our objectives are three-fold. Firstly, we outline the requirements and components of a framework that accommodates hardware diversity. Next, we assess the impact of device heterogeneity on AI inference performance, identifying strategies to optimize outcomes without compromising service quality. Lastly, we shed light on the prevailing challenges and opportunities in this domain, offering insights for both the research community and industry stakeholders.
\end{abstract}

\section{Introduction}
\label{sec:intro}
Over the past decade, Artificial Intelligence (AI) has deeply transformed the landscape of computer technologies, influencing several business verticals, including industrial automation, automotive industry, eXtended Reality (XR), unmanned aerial vehicles (UAVs), and healthcare \cite{zhou2019edge}. Integrating AI capabilities into these segments often involves sophisticated machine learning (ML) algorithms, like deep learning. Such algorithms might necessitate dedicated hardware platforms for both training and inference tasks. The efficient execution of AI inference is crucial for applications demanding strict real-time AI processing and maintaining a specified quality of service (QoS) \cite{itu2030}. The convergence of an increasing demand for high performance and the prominence of \textit{Edge AI} has catalyzed the advancement of cutting-edge AI-ready hardware and software solutions \cite{9905603}. This momentum is evident in the escalated efforts of chipset manufacturers and software developers, who now prioritize customized solutions to harness the full potential of \textit{Edge AI}. For instance, there has been a marked rise in the development of AI accelerator chipsets—such as GPUs, VPUs, and TPUs—explicitly crafted for Edge AI systems and inference tasks \cite{sipola2022artificial}. This growing interest in the industry has naturally spurred the research community to delve deeper, resulting in a surge of research contributions to the domain and making the literature on Edge AI inference expand rapidly.

\begin{figure}[b!]
\vspace{-3mm}
  \centering
  \includegraphics[width=0.95\linewidth]{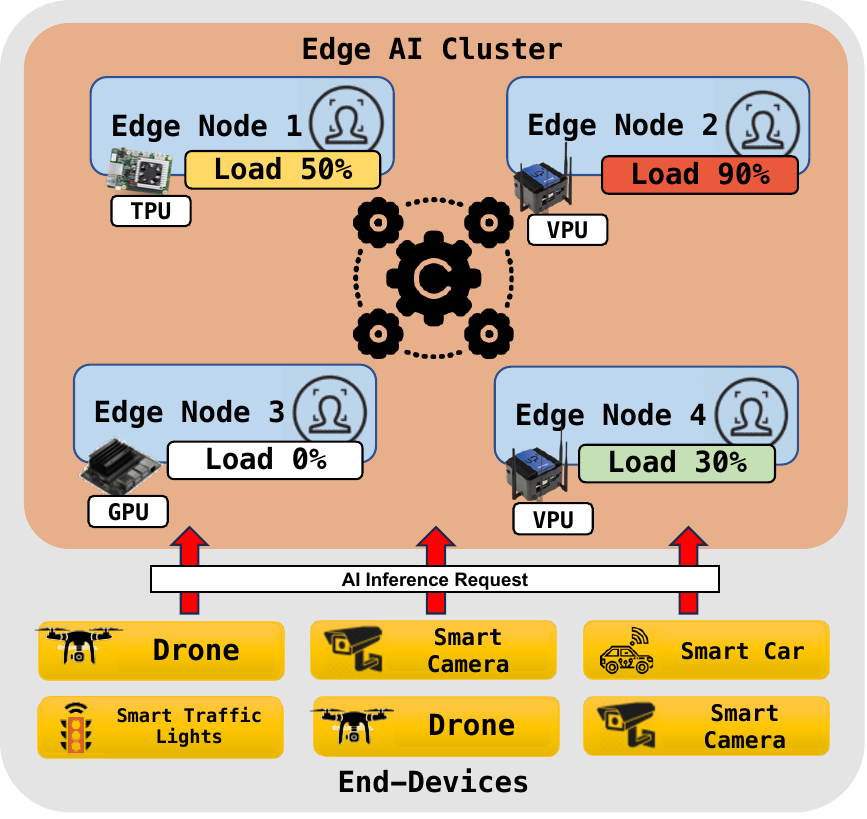}
  \caption{A cluster of edge devices featuring heterogeneous capabilities performs AI inference tasks over data transmitted by multiple end-devices.}
  \label{fig:mainscenario}
\end{figure}

Numerous studies have probed the implications of hardware heterogeneity in single-device systems \cite{jeong2022band}. Others have explored scenarios where AI models are split, and individual AI inference tasks are performed across distributed edge nodes in a multi-node collaborative approach \cite{he2022pyramid}. Our work centers on \textit{single-node inference}. Although both single-node and multi-node collaborative inferences possess their unique advantages and challenges, our preference for the former largely stems from the complications it avoids compared to the model splitting option. These complications encompass unsupported neural network operators due to inconsistencies in hardware and software across different nodes, added overhead in communication and data exchange, and pressing security concerns. Moreover, while many existing solutions lean on theoretical frameworks promising robust optimality \cite{xu2022distributed}, the practicalities and resource limitations of edge nodes in real deployment, exemplified by sophisticated networked systems like vehicular networks \cite{yang2021edge}, drive our emphasis on single-node inference, ensuring its viability across varied deployment contexts.

In this respect, as AI continues to integrate deeper into evolved distributed edge systems, we can anticipate a diverse range of AI accelerator chipsets (e.g., GPU, TPU, VPU) jointly powering future services (Fig. \ref{fig:mainscenario}). This variety imposes additional challenges, encompassing \textit{coordination mechanisms} between devices \cite{9052677} and ensuring that Edge AI inference services are uniformly provisioned and managed across clusters of devices with varied capabilities.
Within such deployments, an end-device seeking Edge AI inference (e.g., applying an object detection algorithm to a video stream) should effortlessly find the desired algorithm within the edge cluster, regardless of the AI accelerator that will execute it. Overcoming this challenge necessitates hardware-agnostic AI service discovery mechanisms. Moreover, the current absence of standardized interoperability among different AI-enabled edge devices, attributed to diverse software tools and APIs, accentuates the need for a unified API to facilitate smooth inter-device communication.
Lastly, it is paramount to ensure optimal resource management in the cluster and the fulfillment of QoS requirements. In this context, QoS entails the entire duration required for an AI request from an end-device to be processed by the Edge AI cluster. This is a demand posed by AI applications. Consequently, offering multiple AI services necessitates strategies for task allocation and orchestration. These strategies should account for computational limits of edge devices, variations in network latency, and specific QoS timings. Additionally, they must consider the unique aspects of the AI application being utilized and how hardware diversity influences AI performance.

Given the complex requirements of these scenarios, the remainder of this paper aims to achieve three primary objectives: \emph{(i)} to introduce the foundational components necessary for a framework to comply with such requirements, \emph{(ii)} to assess the impact of device heterogeneity on AI inference task outcomes, and \emph{(iii)} to spotlight key research and developmental challenges, thereby charting a path forward for both academia and industry.

\section{An Hardware-agnostic AI Inference provisioning}
\label{sec:sectiontwo}

In this section, we first present the hardware resources utilized for developing our proof of concept, and then the main components of the software framework developed to satisfy the requirements mentioned in Section \ref{sec:intro}.

\subsection{Edge AI Cluster: Hardware}
\label{sec:sectionthree}

\begin{figure}[b!]
\centering
  \includegraphics[width=1\columnwidth]{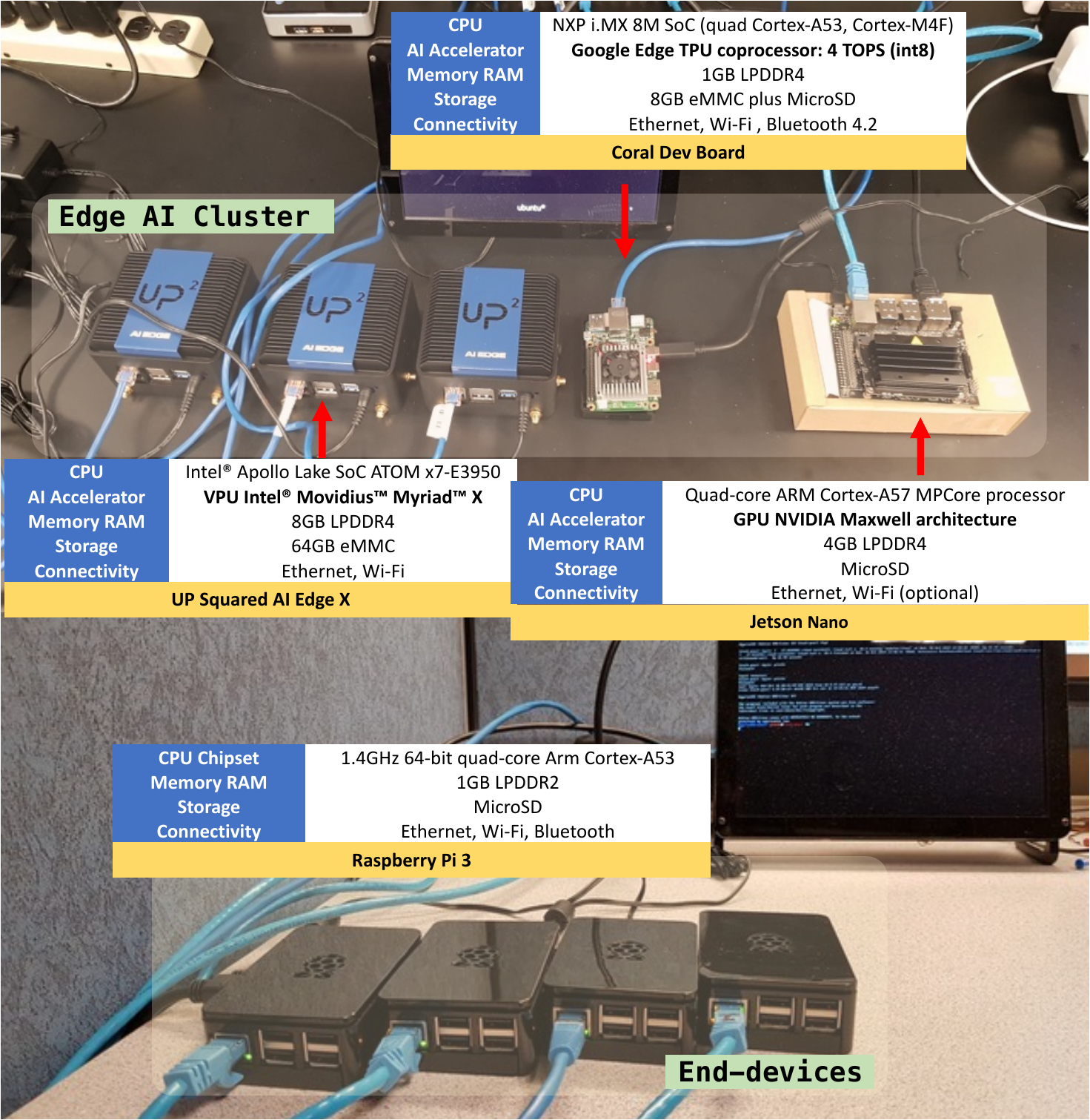}
  \caption{Our testbed includes a set of AI-enabled edge devices (Edge AI Cluster) and multiple end-devices. The figure also presents the detailed list of hardware features of all the devices.}
  \label{fig:testbedall}
\end{figure}

In our experimental testbed setup (Fig. \ref{fig:testbedall}), we introduce edge heterogeneity by using three distinct AI accelerators: \emph{(i)} Intel Movidius Myriad X VPU, \emph{(ii)} Google Edge TPU, and \emph{(iii)} NVIDIA 128-core Maxwell GPU. The three AI accelerator chipsets are respectively hosted in three different Single-Board Computers (SBCs): \emph{(i)} UP Squared AI Edge X, \emph{(ii)} Coral Dev Board, and \emph{(iii)} Jetson Nano. The three edge SBCs have heterogeneous features also from CPU, RAM, and storage perspective. In the role of end-devices, we use four Raspberry Pi 3 Model B.

In the context of our experiments, the devices communicate with each other through a controlled wireless network. Since all of our devices in the Edge AI testbed are placed in the same location, we need to emulate a realistic edge system deployment and so the network latency between devices. 
To determine which distribution to use, we measure latency between a 5G device and a edge server over a 5G commercial network and find the best-fitting distributions to experimental latency values. We observe the best fit of a Stable distribution with parameters of shape 1.6878 and scale 0.0980 to communication latency. The average network latency was found to be 13.405, with a standard deviation of 16.065, indicating the variability in latency measurements.

\subsection{Edge AI Cluster: Software}

\begin{figure}[ht!]
  \centering
  \includegraphics[width=1\linewidth]{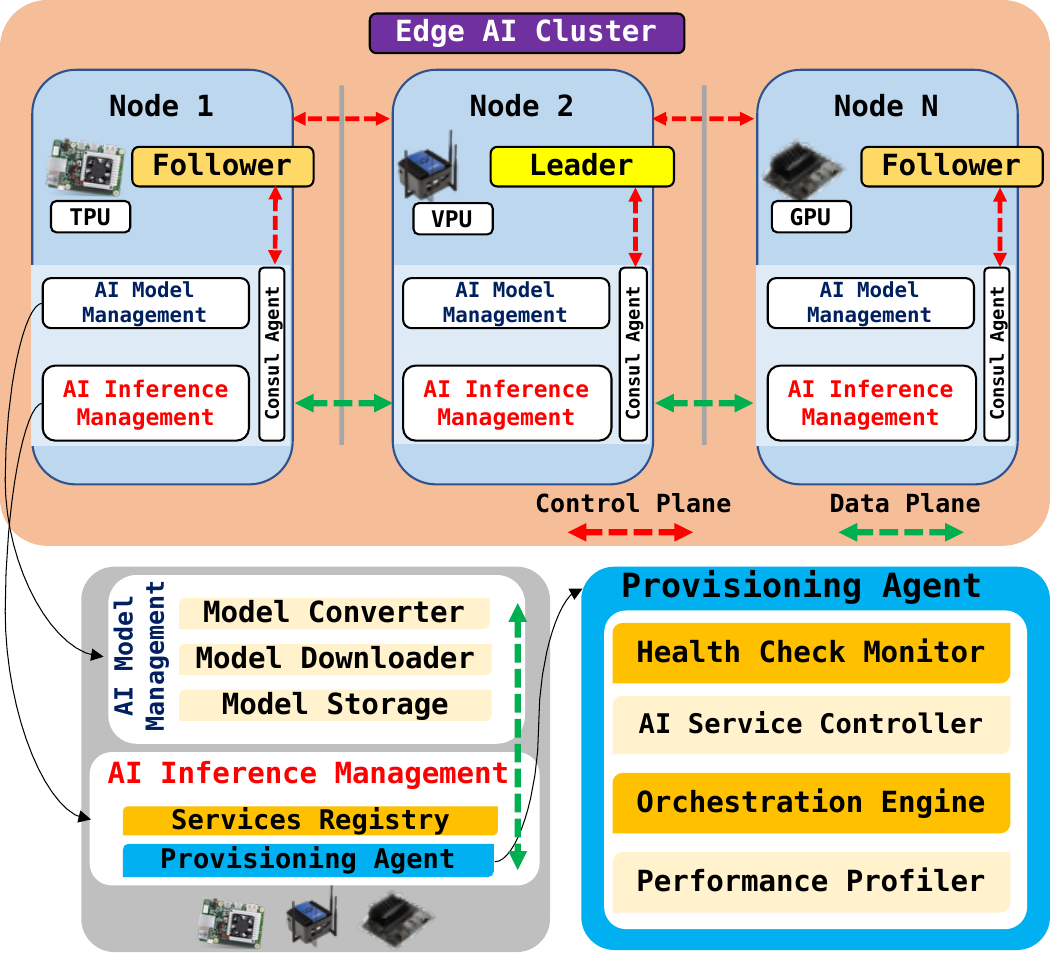}
  \caption{Software architecture and detail of the AI Inference Management component. The Edge AI cluster devices coordinate between each other following a service networking-based approach.}
  \label{fig:softarch}
  \vspace{-3mm}
\end{figure}

The software architecture of each node in the Edge AI Cluster (Fig. \ref{fig:softarch}) is characterized by three main components and several sub-modules. It is designed to meet the requirements of building an abstraction layer that: \emph{(i)} enables interoperability between different AI-enabled devices, \emph{(ii)} allows platform-agnostic service discovery and provisioning of AI inference services, and \emph{(iii)} supports seamless service orchestration and execution migration capabilities.

\textbf{Consul Agent.} Consul \cite{consul} is a versatile service networking solution that ensures secure network connectivity among services deployed in various runtimes. It offers multiple control plane solutions, such as service configuration, service discovery, and service mesh. A service mesh is typically composed of a control plane and a data plane. The control plane is responsible for maintaining a central registry that tracks all services and their corresponding network locations. In our case, the control plane enables communication between the \textit{AI Model Management} and \textit{AI Inference Management} services distributed among the edge devices, and enforces rules and other operational aspects in the service mesh interactions. The control plane uses a consensus protocol for consistency and a gossip protocol to manage membership and broadcast messages to the cluster \cite{consul}. The data plane oversees communication between services according to their specific design (e.g., through REST API). Upon considering an extremely high-frequency gossip interval (0.015ms), we have empirically estimated that the maximum bandwidth required for each Edge AI node to perform all the necessary coordination mechanisms is 7291.7 kbps per node.
The Consul Agent is the core process of Consul and, in our deployment, runs on every cluster node and end-device. End-devices use Consul for AI Inference service discovery. By performing DNS queries, end-devices can easily discover the AI inference services provided by the Edge AI Cluster. For instance, an end-device seeking an object detection (objd) service can use the DNS server directly via name lookups (i.e., \texttt{objd.inference.service.consul}). The query automatically searches for nodes offering the objd service and returns a list of healthy nodes capable of providing that service at that moment.

\textbf{AI Model Management.} This component is responsible for streamlining the process of downloading, converting, and storing various AI models on individual edge nodes. Due to the diverse AI chipsets in our testbed, each hardware unit possesses a unique instruction set architecture, necessitating a platform-specific model compiler to leverage AI-optimized hardware. In simple terms, executing AI inference tasks on these devices requires a distinct AI model format for each platform, which can only be produced after a platform-specific conversion. Transitioning from a model suitable for cloud computing to one compatible with these devices is often a complex and challenging process, as most AI models are not readily available in these newer and less common formats. Additionally, the conversion process must consider the model's data size, hardware processing needs, and overall size and complexity due to the computing limitations of the devices in use. In our scenario, utilizing a particular model across our cluster necessitates converting the model for each target platform \cite{antonini2019resource}. The \textit{Model Converter} was created to simplify this conversion process by automating various steps and intelligently determining specific conversion parameters. Our software implementation is fully compatible with commonly used models.

\textbf{AI Inference Management.} This component leverages not only the specific features of Consul but also incorporates custom implementations developed in the context of this work. It comprises the \textit{AI Services Registry} and the \textit{Provisioning Agent}, which include several sub-components. 
The AI Services Registry maintains a list of locally available AI inference services on each edge node. With Consul, services can be defined in a configuration file or added at runtime through dedicated HTTP APIs. In our setup, AI inference services are exposed to the AI Service Registry via the \textit{Provisioning Agent}, which uses an underlying micro-web framework and a REST API for managing the AI inference services lifecycle tasks (e.g., activation and deactivation of the service). The \textit{AI Service Controller} serves as the primary interface for the underlying AI Accelerator, such as TPU, GPU, or VPU. It oversees the entire lifecycle of an inference process, tailored to the specifics and requirements of both the underlying hardware and software platforms. This encompasses preparing AI models and data for inference, initiating and executing the inference process on the AI accelerator, processing the results as necessary, addressing hardware-specific issues, and deallocating resources after the inference is complete. The implementation of the \textit{AI Service Controller} is contingent upon the particular AI Accelerator embedded in each edge device. This means, for instance, that the Jetson Nano implementation would vary from those of the Coral Dev Board and UP Squared AI Edge X.
Optimal resource allocation is crucial for successfully fulfilling the QoS demands of AI-processed applications in most use cases mentioned in the introduction. Given the hardware constraints of the platforms, it is essential to: \emph{(i)} profile the performance of individual AI inference instances, \emph{(ii)} monitor the overall system performance of each edge platform, and \emph{(iii)} monitor the network performance (i.e., latency) of each edge platform concerning the end-devices to be served and the other edge nodes. The \textit{Performance Profiler} and \textit{Health Checks} serve as the foundation for allocating AI inference execution to the healthiest cluster node or offloading it to a healthier node if QoS can no longer be satisfied.
These two components address the limited resources of edge devices heuristically by monitoring key metrics, logs, and traces to provide an overview of each device's health status and its running services. Defining \textit{Health Checks} allows for determining when nodes and services are considered healthy or not, and specific thresholds can be set for the monitored metrics increasing system resiliency. In our deployment, we defined multiple health checks to outline real-time device status concerning AI inference services, edge platform, and network. Platforms and services health can be set to three possible states: \textit{pass}, \textit{warn}, and \textit{critical}. When a node's system-level health check is set to \textit{critical}, the device is temporarily considered incapable of processing additional AI inference instances. Similarly, if an AI inference service's health check is set to \textit{critical}, it indicates that the executing device is so loaded that it cannot satisfy the service's inference latency requirement. In such cases, the \textit{Orchestration Engine} improves device resource utilization and performance of instances with violated QoS by offloading the execution of AI inference services to a healthier device in near real-time. The engine relies on local information provided by the \textit{Performance Profiler} and utilizes the \textit{AI Service Controller} for orchestrating and offloading operations.
Platform and application health monitoring parameters vary in granularity depending on underlying hardware capabilities. As platform performance metrics, such as CPU utilization (\%) and RAM utilization (\%), may impact inference application quantitative performance (e.g., latency, root-mean-square error), it is crucial to empirically understand how the heterogeneous structure of our edge cluster directly affects inference performance. In the upcoming section, we first provide additional details on how edge nodes are selected for AI Inference provisioning and offloading. Subsequently, in Section \ref{sec:sectionfour}, we explore this aspect in greater detail through a series of experiments and describe how the \textit{Provisioning Agent} modules use these empirical insights to design tasks' offloading orchestration rules.

\section{Node Selection for AI Inference Provisioning and Offloading}
\label{sec:section_new}

The \textit{Performance Profiler} of each device computes in real-time the inference latency of each object detection (OD) running instance, denoted as \( \textit{InfLat}_i \), and the average inference latency, \( \textit{AvgInfLat} \), over all OD instances. Mathematically, \( \textit{InfLat}_i \) represents the time taken to process the \( i^{th} \) OD instance, and \( \textit{AvgInfLat} \) is given by:
\begin{equation}
\textit{AvgInfLat} = \frac{\sum_{i=1}^{n} \textit{InfLat}_i}{n}
\end{equation}
It also keeps track of the demanded QoS by each AI application and transfers all this information to the \textit{Health Checks} component. This component then uses the following criteria to evaluate the state of each application and the overall system based on their respective overall latency:
\[
\textit{State}(L) = 
\left\{
	\begin{array}{ll}
		\text{pass} & \text{if } L < 75\%\text{QoS} \\
		\text{warning} & \text{if } 75\%\text{QoS} < L < 90\%\text{QoS} \\
		\text{critical} & \text{if } L > 90\%\text{QoS}
	\end{array}
\right.
\]
where \( L \) is \(\textit{InfLat}_i\) for individual applications and \(\textit{AvgInfLat}\) for the overall system.
A percentage of QoS represents the proportion of the maximum allowable time taken to process an AI inference request. For instance, 75\%QoS means the process is completed in 75\% of the maximum allowable time. The \textit{Orchestration Engine} of each device processing AI inference tasks operates according to the inputs provided by the \textit{Performance Profiler} and the status of the \textit{Health Checks}, with the aim of promptly acting in case of QoS requirements violation.
By following this heuristic approach, we can ensure that as soon as the QoS requirements cannot be met---as a result of overload caused by concurrent processing tasks or increased network latency---either the application or system health check will be set to \textit{critical}. While the entire node is in \textit{critical} state, it is temporarily unreachable for processing new incoming requests and the execution of the OD instance with the highest \textit{InfLat} is offloaded to a healthier node of the cluster. If it is the QoS of a single application that is violated, then that particular application will be offloaded to a healthier node of the cluster. The seamless migration between nodes is performed by the \textit{Orchestration Engine}, which transfers specific metadata that includes information on the AI model to be used and the video streaming source. The choice of the node to which offload the task execution takes into account the network latency inter-edge devices and end-to-edge devices. Each edge device keeps track of the network latency status among all the different nodes of the system  through a \textit{network latency matrix} (NLM). The ranking of the network latency status between two nodes (e.g., \textit{edge node X} $\leftrightarrow$ \textit{end-device Y} or \textit{edge node X} $\leftrightarrow$ \textit{edge node Z}) is set to \textit{pass}, \textit{warning}, or \textit{critical} not only considering the latest instantaneous value of the latency, but also its most recent time-series evolution.
In particular, in order to rank nodes based on network latency while taking into account both recent and longer-term performance, a methodology employing Exponential Moving Averages (EMAs) for 1-minute, 5-minute, and 15-minute intervals is utilized. To quantify this approach, we denote the Exponential Moving Averages at 1-minute, 5-minute, and 15-minute intervals as \( EMA_{1m} \), \( EMA_{5m} \), and \( EMA_{15m} \) respectively. The composite score, \( S \), for a node can then be expressed as:
\[ S = w_{1m} \cdot EMA_{1m} + w_{5m} \cdot EMA_{5m} + w_{15m} \cdot EMA_{15m} \]
Where \( w_{1m} \), \( w_{5m} \), and \( w_{15m} \) are the respective weights for the EMAs at the 1-minute, 5-minute, and 15-minute intervals. The choice of weights will determine the importance of each time interval in the final score, with lower weights for shorter intervals emphasizing longer-term trends. By combining these EMAs using this weighted sum, a composite score for each node is generated, which emphasizes longer-term trends and helps mitigate short-term fluctuations.
Nodes are then ranked according to their composite scores, ensuring efficient resource allocation and optimal system performance in varying network conditions.
It is worth noting that while there are alternative methods to EMA for time-series analysis \cite{ferreira2023forecasting}, we opted for EMA at this stage due to its straightforward implementation and minimal resource overhead, ensuring the methodology does not significantly add to the resource demand on each node.
\begin{figure}
         \centering
         \includegraphics[width=1\linewidth]{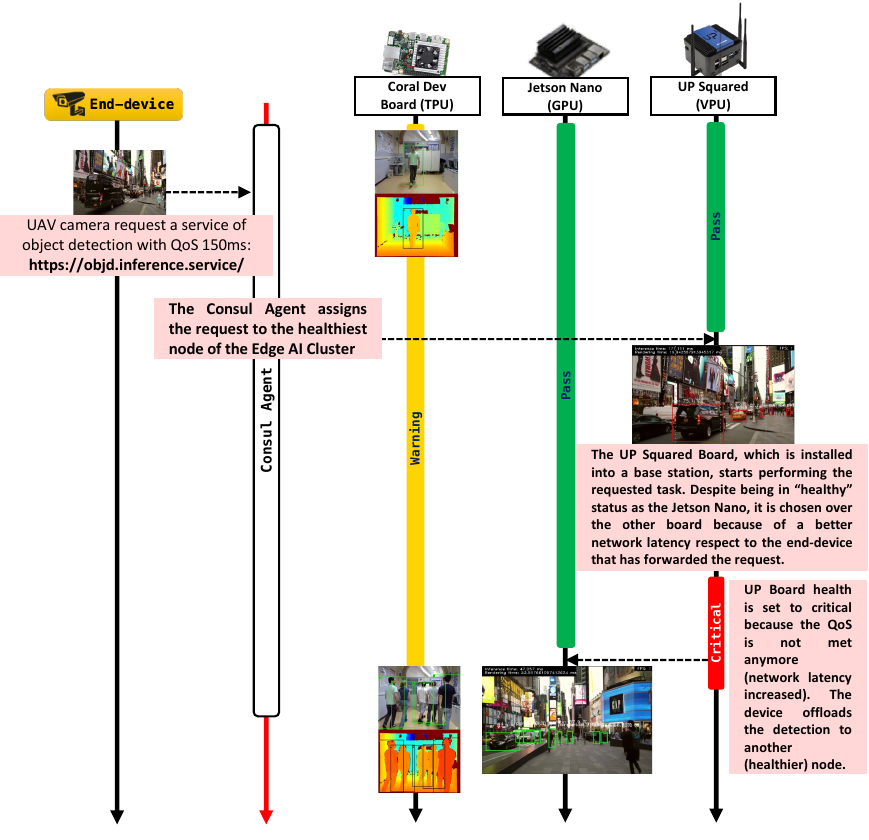}
        \caption{The process of discovery, provisioning, and orchestration of an Edge AI-based object detection service.}
        \label{fig:timediagram}
          \vspace{-3mm}
\end{figure}

 Fig. \ref{fig:timediagram} illustrates the process of discovery, provisioning, and orchestration offloading of an Edge AI-based object detection service in a real-deployment environment. The system allocates the service execution to the healthiest edge device of the cluster (in the example, the UP Squared). As soon as the edge device can no longer guarantee the requested QoS, the AI Inference execution is seamlessly migrated to another edge node (regardless of the underlying AI chipset in use). Algorithm \ref{alg:algorithm} provides a step-by-step overview of the process, highlighting how edge nodes are initially selected for provisioning and how the execution offload is performed. Please note that the text in blue refers to the mechanism of node selection for AI Inference provisioning, while the text in green to the AI Inference execution offload.

\begin{algorithm}
\caption{Select node for AI Inference provisioning and AI Inference execution offload}
\begin{algorithmic}[1]
\scriptsize
\Require Set of edge nodes (\textit{edgeNodes}), end-device (\textit{endDevice}), NLM (\textit{NetworkLatencyMatrix}), QoS requirements
\Ensure Selected healthy edge node (\textit{selectedNode})
\State Preload AI models in \textit{edgeNodes}
\State Initialize \textit{selectedNode} to null
\color{ao}
\Function{assignNode}{\textit{edgeNodes}, \textit{endDevice}, \textit{latencyMatrix}}
    \State Initialize \textit{minLatency} to max(\textit{NetworkLatencyMatrix})
    \State Initialize \textit{healthyNode} to null
    \For{each \textit{edgeNode} in \textit{edgeNodes}}
        \If{\textit{edgeNode} is healthy and the latency between \textit{edgeNode} and \textit{endDevice} is less than \textit{minLatency}}
        \State Update \textit{minLatency} with the latency between \textit{edgeNode} and \textit{endDevice}
            \State Update \textit{healthyNode} with the current \textit{edgeNode}
        \EndIf
    \EndFor
    \State Assign the AI inference task to \textit{edgeNode}
\EndFunction

\State \textit{selectedNode} $\leftarrow$ \Call{assignNode}{\textit{edgeNodes}, \textit{endDevice}, \textit{NetworkLatencyMatrix}}
\color{ao(english)}
\While{true}
    \State Continuously monitor the health status and QoS requirements for \textit{selectedNode}
    \If{\textit{selectedNode} cannot guarantee required QoS for handled applications}
        \State Temporarily make \textit{selectedNode} unreachable for new requests
        \State Find another healthy node as described in the \textsc{assignNode} function
        \If{a new healthy node is found}
            \State Perform a seamless migration of the affected AI inference task to the new healthy node
        \EndIf
    \EndIf
\EndWhile
\\
\Return \textit{selectedNode}

\end{algorithmic}

\label{alg:algorithm}

\end{algorithm}

\section{Heterogeneity Impact on the AI Inference Performance}
\label{sec:sectionfour}
\begin{figure*}[ht!]
  \centering
  \includegraphics[width=1\linewidth]{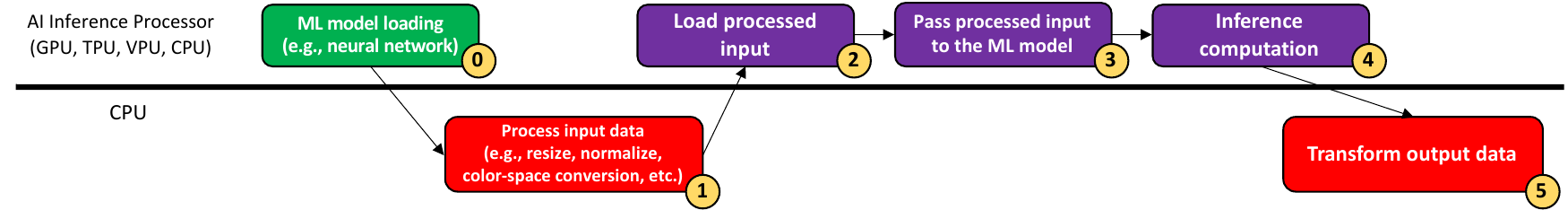}
  \caption{AI Inference Lifecycle. The execution of the different \textit{steps} are executed locally with this order, regardless of the AI chipset in use, in all the edge devices performing inference tasks.}
  \label{fig:inflifecycle}
  \vspace{-3mm}
\end{figure*}

For understanding how hardware heterogeneity impacts the performance of AI inference services, we should first describe what is the typical lifecycle of this kind of instances. Regardless of the edge platform in which the task is executed and the algorithm in use, the overall AI inference processing sequentially occurs between CPU and AI Accelerator (i.e., GPU, TPU, VPU) in sequence as shown in Fig. \ref{fig:inflifecycle}.

The AI model loading is typically a resource-demanding task (especially for devices with limited CPU and RAM resources), therefore it is a good practice to design AI inference applications in a way that \textit{Step 0} is executed only once. Taking this for granted (our implementation complies with this requirement), it is possible to observe how input data is \textit{inferenced} only after a sequence of processing stages that are, in most of the cases, bounced between the CPU and the AI accelerator chipset in use. There is however the exception of the case in which also the inference-related steps are demanded to the CPU. However, this scenario is for now out of the scope of this study.

In our experiments, the AI application executed by the edge devices encompasses the use of the MobileNetV1 algorithm performed over a video content streamed by each of the end-devices.
The model input is a blob that consists of a single image of (1, 300, 300, 3), where the tuple represents batch size, image width, image height, color channels. The choice of a specific image width and height is extremely important. When the base model is first trained with such parameters and then converted for being executed in each targeted platform, the images feeding the model's input at runtime must comply with the choice of such parameters. 

The scope of our empirical analysis is threefold:

\begin{enumerate}
  \item We aim to assess the scaling capabilities of the different AI inference accelerators when processing multiple instances in parallel.
  \item We want to estimate what is the impact produced by pre-processing (\textit{Step 1}) and post-processing (\textit{Step 5}) tasks in the overall inference latency time.
  \item We want to understand to what extent changing the input parameters of the AI-processed application would affect the inference latency performance. 
\end{enumerate}

To accomplish this, we defined two additional experimental scenario requirements. First, we assume the edge devices to serve a growing number of OD instances. Second, we provision the AI model with images having a input frame size higher respect to the frame size used for training the model (i.e., 300x300). This assumption in made to comply with the unpredictability of the application characteristics (video streaming in this specific case) that must be handled. As for the choice of the QoS requirement, we selected an arbitrary value of 150ms, although this can vary from case to case and be therefore more or less stringent \cite{cartas2019reality}. In this study, we overlook the qualitative performance (e.g., mean average precision) of the OD, defined as the ratio between the number of correctly recognized objects and that of total objects in a video frame \cite{liu2018edge}. 
\begin{figure}[]
         \centering
         \includegraphics[width=\columnwidth]{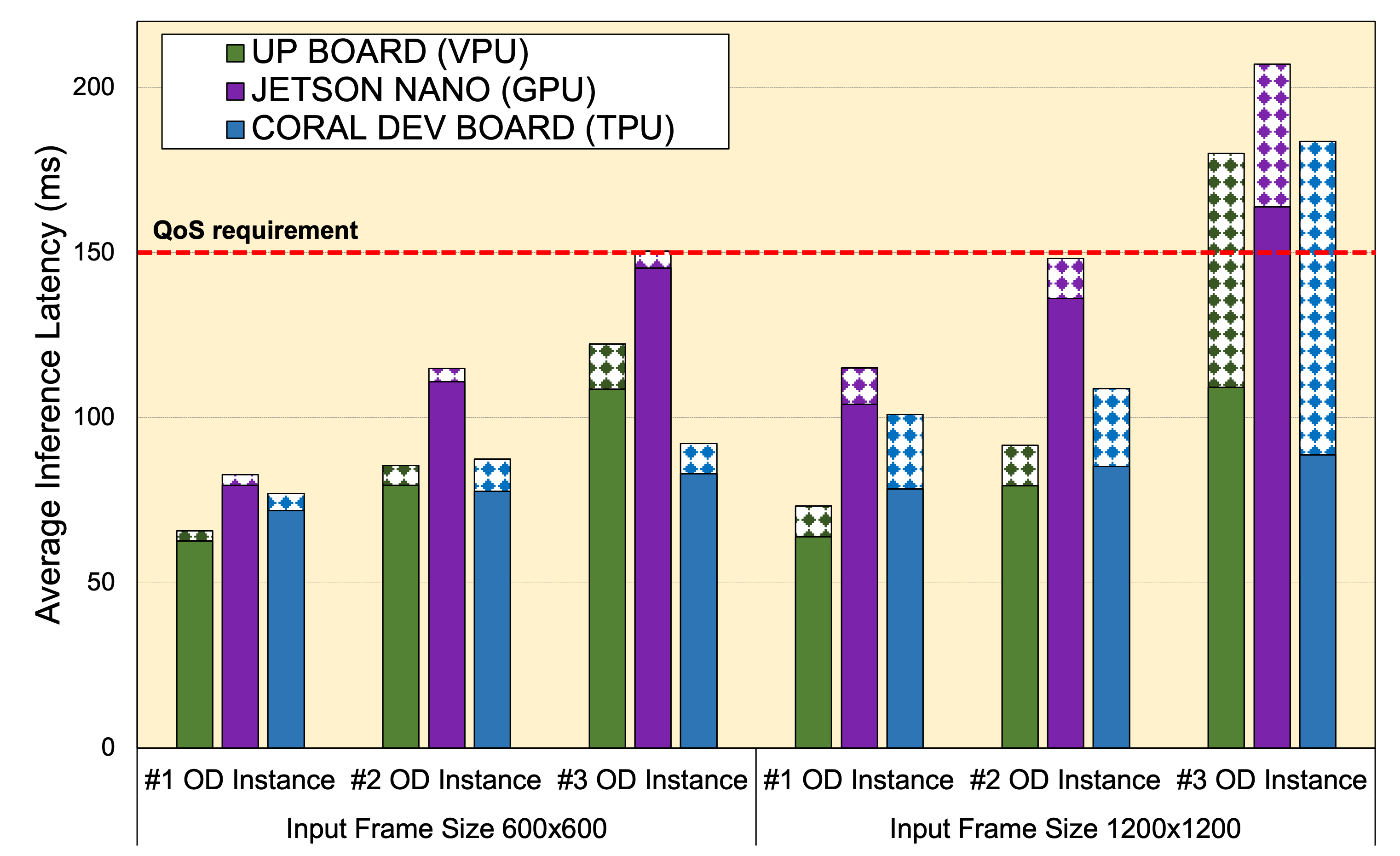}
        \caption{Average AI inference latency breakdown: The solid fill represents the time required for performing Steps 2-4, while the pattern fill indicates the time needed for executing Steps 1 and 5 of the AI Inference lifecycle.}
        \label{fig:fig4b}
        \vspace{-6mm}
\end{figure}

In the context of our experiments, Fig. \ref{fig:fig4b} displays the breakdown of AI inference latency over two distinct components. Specifically, the graphs illustrate the component executed by the CPU, encompassing all frame processing-related tasks (Steps 1 and 5 in Fig. \ref{fig:inflifecycle}). We will refer to this as the CPU component. In contrast, the AI Accelerator component encompasses AI inference processing-related computations (Steps 2, 3, and 4 in Fig. \ref{fig:inflifecycle}).
For our analysis, we evaluated performance metrics for each device under test. These metrics are based on varying the number of OD instances running simultaneously, alongside two different input frame sizes (600x600 and 1200x1200) of the content streamed by the end-devices.
Upon observation, it is evident that for the same input frame size and number of OD instances, different boards yield varied AI Accelerator and CPU performance metrics. As a specific example, with an input frame size of 600x600 and a single OD instance, the UP Squared's AI Accelerator processes the task in roughly 62.65 ms. In comparison, the Jetson Nano requires about 79.59 ms, and the Coral Dev Board finishes in an approximate 71.81 ms. There's also a discernible variance in the performance of the CPU component across these boards.
An increase in the number of OD instances predictably augments the processing duration for both the AI Accelerator and CPU components across all devices. Moreover, enlarging the input frame size amplifies the processing duration. Notably, this augmentation appears more pronounced for the CPU component, suggesting that the CPU, relative to the AI accelerator, is more susceptible to escalations in input frame size.
In terms of scalability under increasing computational demands, the system's overall latency inflates by about 45-60\% on average for all three scenarios when the frame size is enhanced. Analyzing individual devices, the UP Squared consistently exhibits a latency growth trend comparable to the general increment, with an average escalation of 40-50\% as the frame size doubles. The Jetson Nano device, on the other hand, displays varied responsiveness to frame size increases. The rise in latency ranges between 35-65\%. The most substantial increase is observed during the three concurrent running instances (\#3 OD Instance). Lastly, the Coral Dev Board is notably influenced by enlargements in frame size. Its latency inflates on average by 50-70\%. This device, similarly to the Jetson Nano, is particularly sensitive during the execution of three concurrent instances.
These findings underscore the intricate interplay between the input parameters of AI-processed applications and the resultant AI inference latency. The evident heterogeneity among devices further emphasizes this point, as each device responds uniquely to identical workloads, underscoring divergent performance dynamics in both the AI Accelerator and CPU components. The disparities observed accentuate the necessity of introducing nuanced node allocation strategies tailored to these heterogeneous device behaviors.
For example, we envision the need for node allocation strategies that take into account the different performance characteristics of CPU (pre- and post-processing capabilities) and AI accelerators (inference capabilities) in the Edge AI nodes. This could enable a more fine-grained allocation of tasks, considering the unique performance features of different devices (such as GPUs, VPUs, and TPUs) and their ability to handle multiple AI inference tasks concurrently. To follow this approach, the \textit{Performance Profiler} may assign weights to each Edge AI node: \( W_{\text{cpu}} \) for the CPU Weight, representing the node's CPU capabilities for handling pre- and post-processing tasks (Steps 1 and 5); \( W_{\text{ai}} \) for the AI Accelerator Weight, representing the node's AI accelerator capabilities for handling inference tasks (Steps 2-4); and \( W_{\text{nl}} \) for the Network Latency Weight, factored in when allocating tasks following the NLM approach.

The combined weight, \( W_{\text{combined}} \), for each Edge AI node is given by: 
\[ W_{\text{combined}} = \alpha \times W_{\text{cpu}} + \beta \times W_{\text{ai}} + \gamma \times W_{\text{nl}} \]
where \( \alpha \), \( \beta \), and \( \gamma \) are assigned factors for the respective weights.
The \( W_{\text{combined}} \) is updated whenever the execution of an application is scaled up or down, reflecting the current workload and resource availability of each device. Tasks (video streams) are assigned to nodes based on their combined weights and application requirements.
For new upcoming tasks to be processed, the framework parses the workload features of the incoming request (e.g., resolution for the video streaming and OD case) and allocates the execution of the new task based on the pre- and post-processing requirements, as well as inference requirements needed to handle the request. The system dynamically adapts to changes in the workload by updating the \( W_{\text{combined}} \) of each Edge AI node, ensuring that the most suitable nodes are selected for the current system state.

Addressing these intricate challenges and effectively modeling realistic Edge AI environments is undoubtedly non-trivial. We delve deeper into this complexity and discuss potential solutions in Section \ref{sec:sectionfive}.

\section{Opportunities and Challenges}
\label{sec:sectionfive}

The potential to deploy systems that enable interoperability between heterogeneous AI-enabled devices offers significant advantages, considering the growing prevalence of such devices and the requirements of future networks. Numerous research opportunities and challenges still need to be addressed. This section aims to introduce a non-exhaustive set of the most prominent R\&D efforts required to further enhance the technological landscape in this area. 
Our focus will be on system architecture and deployment aspects. Readers may also notice that the identified challenges share a common theme, fitting within the realm of orchestration and management of distributed Edge AI inference services.
\newline
\textbf{Balancing Load and Network Latency in AI-Enabled Heterogeneous Edge Networks.} The computing environments analyzed in this work face the challenge of balancing load and network latency while adapting to the heterogeneous AI hardware landscape. Different tasks have varying computational requirements, and some tasks are better suited for specific hardware types. For instance, GPUs excel at handling tasks with a high degree of parallelism, while TPUs are more efficient for matrix operations. In different scenarios, the priority of minimizing latency or maximizing resource utilization might vary. Real-time applications, like video analytics for public safety, prioritize low latency to ensure timely responses. In contrast, batch-processing applications, such as analyzing historical data, prioritize resource utilization for efficiency. To address these varying requirements and hardware capabilities, a tunable parameter could be introduced in the system that allows adjusting the balance between load and network latency. This parameter could influence the weights assigned to the Edge AI nodes, making the system prioritize either minimizing latency or maximizing resource utilization based on the specific use case. For instance, if low latency is crucial, the tunable parameter could increase the importance of the \textit{Network Latency Weight} in the combined weight calculation, while decreasing the significance of CPU and AI Accelerator Weights. Conversely, if resource utilization is more important, the parameter could increase the importance of CPU and \textit{AI Accelerator Weights}, while decreasing the significance of the \textit{Network Latency Weight}. In future work, exploring the impact of this tunable parameter on the performance of different applications and investigating methods for automatically adjusting it based on the specific use case, performance metrics, or hardware requirements is crucial. Lightweight techniques such as ensemble learning, incremental learning, and real-time or near real-time Pareto-based optimization could be employed to make these adjustments while considering the resource constraints of edge devices considered in this scenario \cite{9806291}. This would enable the system to adapt to various requirements and hardware types, providing more flexibility and better performance across a range of scenarios in the context of AI-enabled heterogeneous edge networks.
\newline
\textbf{Leveraging Intent-Based Networking for Heterogeneous Edge AI Systems.} Hardware and software heterogeneity, along with the presence of diverse AI-enabled services with different requirements, pose significant challenges to current deployments. These factors underscore the necessity for innovative mechanisms to streamline the management of such complex systems, moving away from rigid hard-coded policies towards a more adaptable and flexible approach. Intent-based networking (IBN) offers promising avenues of research for addressing these challenges and enhancing the capabilities of edge computing systems \cite{khan2021intent}. Using high-level intents defined with specification languages like the Autonomic System Specification Language (ASSL \cite{vassev2007assl}) and NEtwork MOdeling (NEMO \cite{tsuzaki2017reactive}), IBN can craft computing- and networking-aware orchestration mechanisms that dynamically adapt to system conditions, network latency, and the unique demands of AI-enabled services. With the rise of Large Language Models (LLMs) \cite{zhao2023survey}, there lies potential in exploiting these advanced AI models to interpret and even generate IBN intents. Such models can understand the intricacies of application requirements and network conditions, enabling a more refined, contextual, and adaptive orchestration of services. Given their capacity to process vast amounts of information and identify patterns, LLMs can be used to optimize the translation of high-level intents into actionable network configurations, reducing manual interventions and enhancing system adaptability. Furthermore, real-time deployments in edge environments often exhibit a dynamic nature with nodes handling diverse applications over varying time spans. IBN's adaptability, enriched by LLMs, can ensure the seamless updating of configurations and policies based on evolving application types and demands. This recursive adaptability ensures sustained optimal performance, even amidst application and service heterogeneity. By synergizing technologies like NEMO, ASSL, and LLMs, IBN can effectively confront the challenges posed by hardware and software heterogeneity, variable network latency, and the diverse necessities of AI-enabled services, heralding a new era of responsive and intelligent edge computing.
\newline
\textbf{Dynamic Code Generation in Heterogeneous Edge AI Systems with Large Language Models.} In Edge AI environments characterized by a mosaic of hardware platforms such as TPUs, GPUs, and VPUs, there arises an intrinsic challenge: how to seamlessly integrate new applications tailored for diverse underlying platforms. Large Language Models (LLMs) present not just a powerful capability in natural language processing but also a promise in dynamic and automatic code generation \cite{poesia2022synchromesh}. By harnessing this unique capability, we can construct custom solutions in real-time that cater to the ever-changing needs of edge applications. However, the current feasibility of such solutions is heavily predicated on the assumption of using LLMs hosted on cloud services through dedicated APIs. The computational demands of LLMs are significant. While there are attempts to run LLMs on more constrained devices such as smartphones, we are still far from a level that could ensure performing code generation locally on constrained edge nodes. In a standalone edge device cluster, the inference time of these models could indeed be prohibitively long, making real-time code generation challenging if not unfeasible. This underpins the essential role of cloud support in actualizing the benefits of LLMs in edge scenarios. In scenarios where application requirements and system needs constantly evolve, LLMs, supported by these dedicated cloud-based APIs, could potentially generate code segments or configuration scripts that bridge software-hardware gaps, optimize ongoing processes, or even manifest novel functionalities tailored for each type of hardware. Such an approach radically redefines flexibility. Instead of relying on pre-defined solutions or configurations that might not fully align with emergent requirements, the system can generate its own new solutions, tailored perfectly for the context through tailored code sandboxing and trustable SOTA (Software over the air) update mechanisms. This potential goes beyond mere reactionary fixes; it encompasses the proactive, intelligent generation of solutions. These capabilities are not just about meeting immediate needs but also about forecasting future challenges based on observed patterns. Coupled with the ability of LLMs to interpret and act on high-level intents in IBN, this offers a unique opportunity to elevate the efficiency, adaptability, and intelligence of heterogeneous Edge AI systems. However, this novel paradigm also introduces its set of challenges. Ensuring the reliability of dynamically generated code \cite{liu2023your}, maintaining security and integrity during real-time adaptations, and verifying the consistency and correctness of such on-the-fly solutions emerge as critical research avenues.
\section{Conclusion}
Designing edge computing systems that can effectively integrate and manage heterogeneous AI-enabled nodes is crucial for unlocking the full potential of AI Inference capabilities across a diverse set of applications and scenarios. By considering the requirements and challenges outlined in this paper, researchers and practitioners can contribute to the ongoing development of innovative solutions that address the complexities of interoperability, performance optimization, and resource management in these systems. This, in turn, will help pave the way for the widespread adoption and successful implementation of AI-enabled edge computing solutions in various industries and use cases.
\bibliographystyle{IEEEtran}
\bibliography{References}

\end{document}